\documentclass[10pt,conference]{IEEEtran}
\usepackage{lineno} % Add line numbers
\IEEEoverridecommandlockouts
\usepackage{cite}
\usepackage{amsmath,amssymb,amsfonts}
\usepackage{algorithmic}
\usepackage{textcomp}
\usepackage{ifthen}
\usepackage{pifont}
\usepackage{ulem}  % Without the 'normalem' option
\usepackage{adjustbox}
\usepackage{setspace} % Package to control line spacing
\usepackage{wrapfig}
\usepackage{amsmath} % for the equation* environment
\usepackage{stfloats}  % Include this in your preamble

\usepackage{multirow}
\usepackage{float}
\usepackage{lscape} % To make the page landscape
\usepackage{tikz} % For drawing custom checkboxes
\usepackage{tabularx} % For tables with automatic text wrapping
\usepackage{caption}  % For better control of table captions
\usepackage{multicol} % For multi-column text
\usepackage{balance}  % For balancing columns (optional)
\definecolor{BlueGray}{RGB}{182,205, 216}
\definecolor{DarkBlueGray}{RGB}{96, 125, 139}
\definecolor{GreenGray}{RGB}{77,107,83}
\definecolor{safagreen}{HTML}{659447}
\newcolumntype{L}[1]{>{\raggedright\let\newline\\\arraybackslash}p{#1}} 
\newcolumntype{C}[1]{>{\centering\let\newline\\\arraybackslash}p{#1}} 
\newcolumntype{R}[1]{>{\raggedleft\let\newline\\\arraybackslash}p{#1}}

\def\BibTeX{{\rm B\kern-.05em{\sc i\kern-.025em b}\kern-.08em
    T\kern-.1667em\lower.7ex\hbox{E}\kern-.125emX}}
    
\newboolean{showcomments}
\setboolean{showcomments}{true} % toggle to show or hide comments
\ifthenelse{\boolean{showcomments}}
  {
   
  }

\newcommand{\framework}{\textit{SAFE-RL} }
\begin{document}

\title{Evaluating Reinforcement Learning Safety and Trustworthiness in Cyber-Physical Systems}
\author{
    \IEEEauthorblockN{Katherine Dearstyne}  
    \IEEEauthorblockA{
    \textit{University of Notre Dame} \\  
    Notre Dame, IN, USA \\  
    kdearsty@nd.edu}  
    \and  
    \IEEEauthorblockN{Pedro (Tony) Alarcon Granadeno,\\Theodore Chambers}  
    \IEEEauthorblockA{ 
    \textit{University of Notre Dame} \\  
    Notre Dame, IN, USA \\  
    \{palarcon, tchambe2\}@nd.edu}  
    \and  
    \IEEEauthorblockN{Jane Cleland-Huang}  
    \IEEEauthorblockA{
    \textit{University of Notre Dame} \\  
    Notre Dame, IN, USA \\  
    JaneHuang@nd.edu}  
}

% \author{Anonymous Authors}
% \author{\IEEEauthorblockN{Katherine Dearstyne, Pedro (Tony) Alarcon Granadeno, Theodore Chambers, Jane Cleland-Huang}
% \IEEEauthorblockA{\textit{Computer Science and Engineering} \\
% \textit{University of Notre Dame}\\
% Notre Dame, IN, USA\\
% \{kdearsty, palarcon, tchambe2, JaneHuang\}@nd.edu}
% }
\maketitle

\begin{abstract}
Cyber-Physical Systems (CPS) often leverage Reinforcement Learning (RL) techniques to adapt dynamically to changing environments and optimize performance. However, it is challenging to construct safety cases for RL components. We therefore propose the  SAFE-RL (Safety and Accountability Framework for Evaluating Reinforcement Learning) for supporting the development, validation, and  safe deployment of RL-based CPS. We adopt a design science approach to construct the framework and demonstrate its use in three RL applications in small Uncrewed Aerial systems (sUAS).
\end{abstract} 
%, and RL-based components may make incorrect decisions when encountering rare or unseen scenarios that they haven’t learned to handle, potentially leading to unpredictable and unsafe behavior. 
%
%%%%%%%%%%%%%%%%%%%%%%%%%%%%

%%%%%%%%%%%%%%%%%%%%%%%%%%%%
\begin{IEEEkeywords}
reinforcement learning, CPS, safety
\end{IEEEkeywords}

\section{Introduction}
\label{sec:introduction}
Reinforcement Learning (RL) has become a key enabler of self-adaptive behaviors in Cyber-Physical Systems (CPS). By adapting to dynamic and uncertain environments, RL enhances system responsiveness but introduces inherent uncertainties which pose challenges to comprehensive safety assurances \cite{SurveyRLSafety}. We therefore present the \framework framework (Safety and Accountability Framework for Evaluating Reinforcement Learning), developed using an iterative design science approach with RL experts. The resulting framework---which encompasses safety goals, risk-related questions, and mechanisms for providing evidence and traceability of safety-critical decisions \cite{DBLP:conf/re/MarczakCzajkaNC23}---offers a systematic way to assess how safety is addressed in the design and verification of RL components in CPS. We demonstrate \framework's use through examples depicting how RL-based control mechanisms
can satisfy stringent safety requirements for real-world deployment of small Uncrewed Aerial systems (sUAS).

\begin{figure}[h]
    \centering
    \includegraphics[width=\columnwidth]{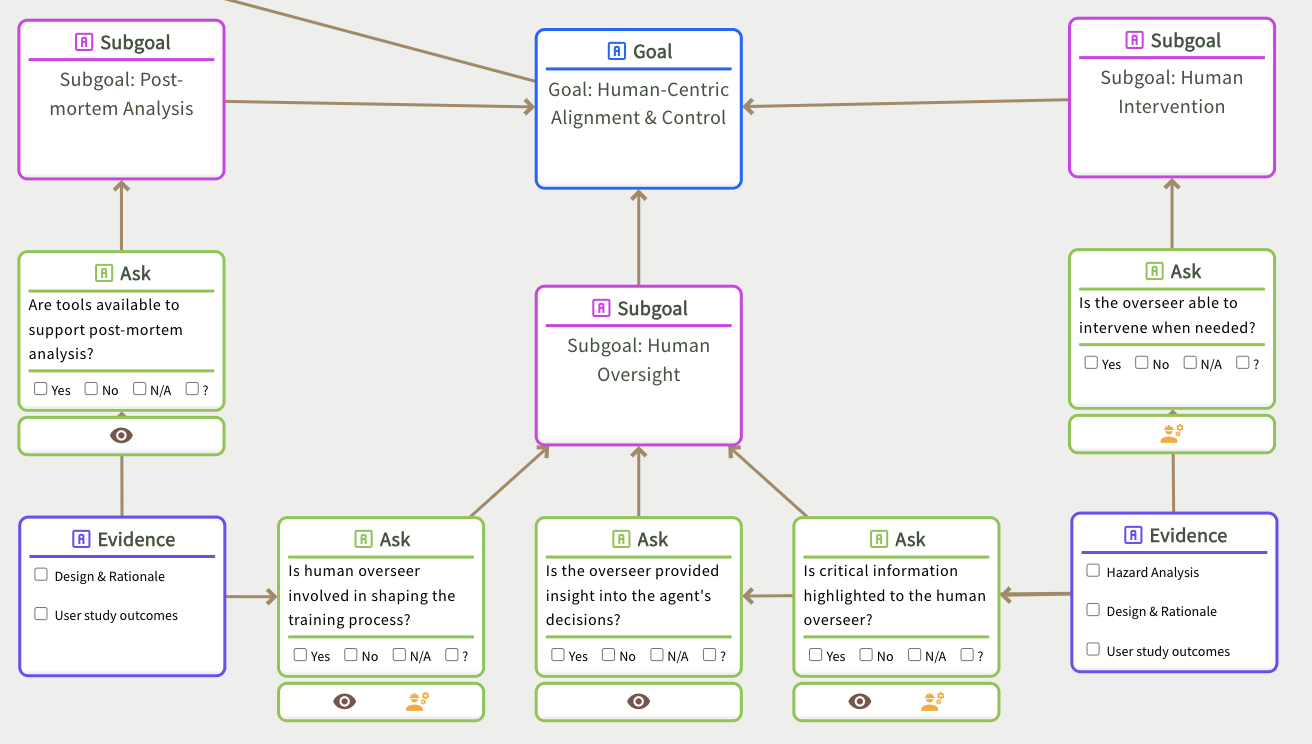}

    \caption{Hierarchical breakdown for Goal 3. Each \texttt{ASK} node shows an icon for its related trustworthiness dimensions.}
    \label{fig:example-goal3}
    \vspace{-1em}
\end{figure}
\raggedbottom

% Developing safe RL systems presents several significant challenges \cite{AmodeiAISafety, SurveyRLSafety}. Key issues include handling novel scenarios \cite{DBLP:journals/corr/abs-2111-09794}, designing reward functions that prioritize safety \cite{gu2022review}, addressing risks introduced by post-deployment learning and adaptation \cite{khetarpal2022towards}, providing formal safety guarantees \cite{RamaswamyTheoryvPractice}, improving transparency and explainability \cite{iyer2018transparency}, and ensuring the reliability of state information in dynamic environments  \cite{dulac2021challenges}. 

\section{\framework Framework }
\label{sec:framework-overview}

% C0 has extensive experience in CPS, while C1-C2 both have expertise in RL. The process is described in Table \ref{tab:design-steps}.

% \begin{table}[H]
% \caption{Description of each design iteration. }
% \label{tab:design-steps}
% \centering
% \footnotesize

% \setlength{\tabcolsep}{6pt} % Adjust column spacing for better appearance
% \renewcommand{\arraystretch}{1.3} % Adjust row height for readability

% \begin{tabular}{p{.2cm} p{7cm}}
%     \hline
%     \midrule
%      {\bf 1} & \textit{C0} designed an initial goal hierarchy—covering Environment Alignment, Safety, and Transparency \& Explainability—along with \texttt{ASK} and evidence nodes.    \\ 
%     \midrule
%     {\bf 2} & \textit{C1 \& C2} reviewed the hierarchy, emphasizing the need for greater specificity in \texttt{ASK} nodes. Their feedback led to refinements and the addition of five \texttt{ASK} nodes, including one for multi-agent training.  \\ 
    
%     \midrule
%     {\bf 3} & \textit{E1} provided further feedback, refining sub-goals and adding three new ones. The high-level goals were revised to  better align with related sub-goals and \texttt{ASK} criteria. Additionally, \texttt{ASK} nodes were refined to explicitly consider robustness.  \\ 
%     \bottomrule
%       \multicolumn{2}{c}{*C=co-author; E=external expert}\\  
% \end{tabular}
% \vspace{-1em}
% \end{table}

\subsection{Framework Hierarchy}
\renewcommand{\arraystretch}{1.4} % Increase row height for better readability
\begin{table}[t] % Use H to force the table to appear precisely here
\footnotesize
\centering

\caption{Descriptions of \framework Goals and Subgoals}
\begin{adjustbox}{width=0.48\textwidth} % Adjusted width for single-column fit
\begin{tabularx}{0.48\textwidth}{|>{\raggedright\arraybackslash}X|}
\hline
% \rowcolor{BlueGray}
\multicolumn{1}{|c|}{\textbf{Goal 1: Real-World Readiness \& Adaptability}} \\

% \multicolumn{1}{|>{\centering\arraybackslash}X|}{\textit{Training scenarios
% align with real-world deployment environment}} \\

\hline
\textbf{Subgoal 1.1.\textit{ Scenario Diversity \& Coverage}}: Emphasizes the inclusion of a diverse and comprehensive set of scenarios during training to improve agent preparedness. \textbf{({\# of \texttt{ASKs:} 3})} \\
\textbf{Subgoal 1.2.\textit{ Design-Environment Alignment}}: Evaluates whether the design choices for the RL agent, such as the algorithm and architecture, are well-suited to the characteristics of the real-world environment.  
\textbf{({\# of \texttt{ASKs:} 1})} \\
\textbf{Subgoal 1.3.\textit{ Resilience \& Adaptation in Real-World Deployment}}: Assesses the system's capability to adapt to real-world conditions and manage unexpected challenges effectively. 
\textbf{({\# of \texttt{ASKs:} 2})} \\
\hline
% \rowcolor{BlueGray}
\multicolumn{1}{|c|}{\textbf{Goal 2: Risk Management}} \\
% \multicolumn{1}{|>{\centering\arraybackslash}X|}{\textit{Emphasizes designing training processes that minimize risky behavior and implement safety measures to mitigate failures during deployment.}} \\
\hline
\textbf{Subgoal 2.1 \textit{Risk-Aware Training}}: Outlines strategies to incorporate during training that proactively minimize downstream risks and enhance system safety. \textbf{({\# of \texttt{ASKs:} 3})} \\
\textbf{Subgoal 2.2 \textit{Proactive Risk Aversion}}: Emphasizes the prevention of anticipated risks by enforcing appropriate constraints. \textbf{({\# of \texttt{ASKs:} 1})} \\
\textbf{Subgoal 2.3 \textit{Real-Time Risk Detection}}: Evaluates the system’s ability to identify and respond to safety risks during runtime. \textbf{({\# of \texttt{ASKs:} 1})} \\
\textbf{Subgoal 2.4 \textit{Automatic Overrides \& Failsafes}}: Ensures that the system includes automatic intervention mechanisms to manage situations where the RL agent may fail or act dangerously. \textbf{({\# of \texttt{ASKs:} 1})} \\
\hline
% \rowcolor{BlueGray}
\multicolumn{1}{|c|}{\textbf{Goal 3: Human-Centric Alignment \& Control}} \\
% \multicolumn{1}{|>{\centering\arraybackslash}X|}{\textit{Focuses on fostering trust between human overseers
% and the RL agent by ensuring transparency and offering
% effective control mechanisms.}} \\
\hline
\textbf{Subgoal 3.1 \textit{Human Intervention}}: Ensures that human operators are able to take control of the RL agent when required. \textbf{({\# of \texttt{ASKs:} 1})} \\
\textbf{Subgoal 3.2 \textit{Human Oversight}}: Enhances transparency by providing the human overseer with crucial information during the system’s runtime, enabling informed decision-making. \textbf{({\# of \texttt{ASKs:} 3})} \\
\textbf{Subgoal 3.3  \textit{Post-Mortem Analysis}}: Ensures that sufficient information is available to analyze RL-driven behavior following any system failures. \textbf{({\# of \texttt{ASKs:} 1})} \\
\hline
\end{tabularx}
\end{adjustbox}
\label{tab:safe_rl_summary}
\vspace{-12pt}
\end{table}
\raggedbottom

The \framework adapts elements of Goal Structured Notation to break down the objective of achieving safe and trustworthy behavior in RL systems into a hierarchy of four node types as shown in Figure \ref{fig:example-goal3}: (i) goals (blue), (ii) subgoals (pink), (iii) question (\texttt{ASK}) nodes (green), and (iv) evidence nodes (purple). The top of the hierarchy defines three overarching \textbf{goals} which are  refined into \textbf{subgoals} as described in Table \ref{tab:safe_rl_summary}. Each subgoal is evaluated through \textbf{\texttt{ASK}} nodes, which prompt stakeholders to critically assess design decisions. Unlike traditional Safety Assurance Cases—which may reinforce bias by affirming claims \cite{DBLP:conf/gi/JohnsonL14}—\texttt{ASK} nodes prompt stakeholders to critically evaluate whether their system meets specified requirements. Rather than serving as a simple checklist, this process requires supporting evidence for each \texttt{ASK}, either by selecting predefined options from \textbf{evidence} nodes or providing custom justifications.
The \texttt{ASK} nodes are also mapped to critical dimensions for trustworthy RL systems identified by Xu et al. \cite{xu_trustworthy_2022}: robustness, safety, and generalizability. Additionally, we emphasize a fourth dimension of transparency, focusing on clarity, interpretability, and auditability to support human oversight and evaluation. Although transparency is interwoven with safety, we highlight it separately to underscore the importance of a human-centric design. These dimensions are represented as icons associated with \texttt{ASK} nodes (see Fig. \ref{fig:example-goal3}). In total, the framework includes 7 \texttt{ASKs} for robustness, 6 for generalizability, 10 for safety, and 4 for transparency.

\subsection{Designing and Applying \framework}
\label{sec:framework-design}
To develop the \framework Framework, we adopted the design science research method \cite{DBLP:conf/cibse/Wieringa16} and conducted three iterations of problem investigation, design, validation, implementation, and evaluation. First, a co-author with CPS experience developed an initial hierarchy comprised of the high-level goals of Environment Alignment, Safety, and Transparency \& Explainability. Then, two co-authors (C1,C2), with
experience in RL for sUAS systems, reviewed the hierarchy, and emphasized the need for greater specificity in \texttt{ASK} nodes. Their feedback led to the addition of five \texttt{ASK} nodes, including one for multi-agent training. Finally, an external RL expert (E1) provided further feedback, refining the high-level goals to  better align with related sub-goals and adjusting \texttt{ASK} nodes to explicitly account for robustness.

% \subsection{\framework Framework Construction}

% \begin{itemize}
%     \item {\bf Step 1:}  A co-author with CPS expertise designed an initial goal hierarchy—covering Environment Alignment, Safety, and Transparency \& Explainability—along with \texttt{ASK} and evidence nodes.

%     \item {\bf Step 2:} Two additional co-authors with RL experience reviewed the hierarchy, emphasizing the need for greater specificity in \texttt{ASK} nodes. Their feedback led to refinements and the addition of five \texttt{ASK} nodes, including one for multi-agent training.

%      \item {\bf Step 3:} An external RL expert provided further feedback, refining sub-goals and adding three new ones. The high-level goals were revised to Real-world Readiness \& Adaptability, Risk Management, and Human-Centric Alignment \& Control, better aligning with related sub-goals and \texttt{ASK} criteria. Additionally, \texttt{ASK} nodes were refined to explicitly consider robustness.

% \end{itemize}

% \subsection{\framework Framework Hierarchy} 

% \input{figures/tables/dimensions}
% \subsection{Demonstration with Use Case Scenarios}
% \label{sec:eval}
C1, C2 and E1 then each  applied \framework to a deep-RL use case associated with their work in this area. The scenarios included: obstacle avoidance and maintaining separation distance between sUAS (C1), battery efficiency and route optimization for sUAS (C2), and command and control of small autonomous ground vehicles (E1). Table \ref{tab:use-case-responses} presents examples of their responses for two different \texttt{ASK}s.

% Combined Table for Goals 1, 2, and 3
\begin{table}[h]
\centering
\setlength{\tabcolsep}{4.8pt}  % Maintain column padding to fit in one column

\caption{Responses for Use Cases for 2 \texttt{ASK}s}
\footnotesize
\begin{tabular}{p{3.1cm}|p{3.1cm}|p{1.8cm}}
\hline
% Goal 1

\multicolumn{3}{p{8.3cm}}{
\centering
\textbf{\texttt{\uline{ASK 5:}}  Does the RL system include mechanisms, such as anomaly detection or uncertainty estimation, to detect when the agent encounters out-of-domain states or adversarial scenarios?
} 
% {\scriptsize \newline \ding{111} \texttt{Yes} \ding{111} \texttt{No} \ding{111} \texttt{N/A} \ding{111} \texttt{?}}
}
\\
\hline
{\small \textbf{C1}:} \texttt{YES} \ding{51} \newline \uline{Explanation:} \raggedright \newline --  calculate the Mahalanobis distance of incoming observations from the training distribution to identify OOD scenarios \newline  -- continuous monitor of state variables and trigger alerts when values exceed predefined thresholds. 
& {\small \textbf{C2}:} \texttt{?} \ding{51} \newline \uline{Explanation:}  \raggedright I don't have a complete answer for this. But, assuming we do have some sort of anomaly detection already onboard, along with a capability to detect faulty paths generated by the RL agents, we can at least detect problematic scenarios. & {\small \textbf{E1}:} \texttt{Yes} \ding{51} \newline \uline{Explanation:} In our use case, we use a system dynamic based alg to detect the adversary\\
\hline
% Goal 3
\multicolumn{3}{p{8.3cm}}{
\centering
\textbf{\texttt{\uline{ASK 16:}} Are critical factors or states that significantly impact behavior and safety highlighted for the human overseer?
} 
% {\scriptsize \newline \ding{111} \texttt{Yes} \ding{111} \texttt{No} \ding{111} \texttt{N/A} \ding{111} \texttt{?}}
}
\\
\hline
{\small \textbf{C1}:} \texttt{YES} \ding{51} \newline  \raggedright The system will provide a GUI with: \newline -- proposed paths for each agent. \newline -- minimum separation distances for each drone at each epoch. \newline -- events/RL errors. \newline -- indicator if the RL agent is stuck in local minima and needs a fail-safe transition. & {\small \textbf{C2}:} \texttt{YES} \ding{51} \newline \raggedright The human overseer will see potential paths chosen by RL agents, along with relevant statistical information (confidence, risk of dynamic intrusions, etc.) to understand each agent's state. RL agents can learn to broadcast alerts based on their environmental understanding. & {\small \textbf{E1}:} \texttt{N/A} \ding{55}\newline \raggedright No human overseer involved in the deployment outside of those training it \\
\end{tabular}
\label{tab:use-case-responses}
\vspace{-12pt}
\end{table}

% \subsection{Preliminary Evaluation through Interview with External Participants}
% As a preliminary evaluation of the \framework Framework, we conducted interviews with two external experts (E2, E3) experienced in deep RL for CPS. Each expert was presented with the populated
% SAFE-RL hierarchy based on Use Case 1, and asked to examine it whilst using a
% think-aloud protocol. 

%  Both acknowledged the framework filled a gap in their current practices, with E3 stating it was "more detailed than anything [he] had encountered" and could increase the thoroughness of his deployment process. E2 noted that it was a "good way to conceptually plan out a project before any physical work is done". However, experts did have some recommendations for expanding the tree. E2 stressed the importance of the real-world adaptability goal, suggesting additional nodes for noise handling and sensor-specific considerations. E3 believed that a larger portion of the tree should be dedicated to challenges related to simulation to real-world differences.
%\input{4_related-work.tex}
% \input{5_threats.tex}
%\input{6_future_work}  %Lets merge this into conclusions.
\section{Conclusions} 
\label{sec:conclusions}
Several approaches, such as the AI Safety and Security Checklist \cite{checklist1}, the LivePerson AI System Safety Checklist \cite{checklist2}, and AI Safety Standards such as ISO/IEC 23894:2023, provide processes for general AI Safety, without specifically addressing RL challenges. 

To address this gap, we presented the \framework Framework,  a Q\&A-based approach for evaluating RL system safety. While \framework represents a significant step toward safer and more accountable AI systems, future work will focus on enhancing guidance, incorporating additional safety dimensions, and adapting the framework to diverse AI applications to support a wider range of practitioners.

% This addition could help practitioners avoid common pitfalls and ensure that critical aspects are not overlooked. These and other changes will require extensions to the navigability and usability of our supporting tool.  

% Lastly, we aim to extend the applicability of our ASK-based framework to a broader range of AI systems, such as those used in computer vision. Given the growing need to assess the trustworthiness of diverse AI systems beyond reinforcement learning, this expansion could provide valuable insights and guidance for a wider community of AI practitioners.

%%%%%%%%%%%%%%%%%%%%%%%%%%%%
\section*{Acknowledgment}
The work described in this paper is partially funded by USA
National Science Foundation Grants \#2131515 and \#2103010.
% This work was partially funded by the US National Science Foundation under grant CNS:1931962. We thank undergraduate students list them all here ... 
%%%%%%%%%%%%%%%%%%%%%%%%%%%%

% \section{Appendix} 
% \label{sec:appendix}
% In this appendix we provide models of all three \framework goals and their subtrees in Figure \ref{fig:goals-zoomed} and data from the Use-Case presented to the Experts for assessment in Tables \ref{tab:goal1-res}-\ref{tab:goal3-res}.
% \clearpage
% \newpage
% \input{7_appendix}

%\jch{I'll add more references as we are allowed up to 2 pages!!}
%%%%%%%%%%%%%%%%%%%%%%%%%%%%
% REFERENCES
%\balance{}
\bibliographystyle{IEEEtran}
\bibliography{bibby}

%%%%%%%%%%%%%%%%%%%%%%%%%%%%
% End doc
%%%%%%%%%%%%%%%%%%%%%%%%%%%%
\end{document}